\documentclass[fleqn,10pt]{article}
\usepackage[utf8]{inputenc}
\usepackage[T1]{fontenc}
\usepackage[english]{babel}
\usepackage[letterpaper,margin=1in]{geometry}

\usepackage{amsmath,amsfonts,amssymb}
\usepackage{graphicx}
\usepackage{xcolor}
\usepackage{booktabs}
\usepackage{caption}
\usepackage{authblk}
\usepackage{url}
\usepackage{float}
\usepackage[colorlinks=true,allcolors=blue]{hyperref}
\usepackage[superscript,biblabel,nomove]{cite}
\title{A Synthetically-accessible Universe of Chemically Recyclable Polymers}

\author[1]{Anagha Savit}
\author[1]{Wei Xiong}
\author[1]{Harikrishna Sahu}
\author[1,2]{Shivank S. Shukla}
\author[3]{Will R. Gutekunst}
\author[1,*]{Rampi Ramprasad}
\affil[1]{School of Materials Science and Engineering, Georgia Institute of Technology, Atlanta, GA 30332, USA}
\affil[2]{Matmerize Inc., Atlanta, GA 30332, USA}
\affil[3]{School of Chemistry and Biochemistry, Georgia Institute of Technology, Atlanta, GA 30332, USA}
\affil[*]{Correspondence: rampi.ramprasad@mse.gatech.edu}
\date{}
\begin{document}

\maketitle

\begin{abstract}
Polymers synthesized via ring-opening polymerization (ROP) of cyclic monomers represent an important class of materials due to their chemical recyclability and possible insertion in several critical applications. We present a dataset of 1 million synthetically realizable ROP polymer structures generated through a combination of Virtual Forward Synthesis (VFS) and polymer expert language models and qualified by stringent chemical heuristics. VFS is used to generate ROP polymers by applying known reactions to existing monomers. The polymer foundation models polyBART and POLYT5 further enable the generation of ROP candidates, with polyBART exploring its learned latent space and POLYT5 producing candidates via sequence-to-sequence generation. The resulting ROP polymers are subjected to robust filtering criteria to ensure novelty, validity and overall data quality through a combination of automated validation pipelines and a comprehensive set of chemist-informed heuristic rules introduced in this work for the first time. We hope that this dataset will serve as a valuable resource for downstream sustainable applications.
\end{abstract}

\thispagestyle{empty}

\section*{Background \& Summary}

Polymers are a fundamental component of everyday life owing to their versatile nature and durability. However, this durability also poses a key challenge. Polymers are often difficult to recycle owing to their inherent chemical and thermodynamic stability. This has resulted in a huge buildup of plastic waste at the end of their life cycle \cite{key1,key2}. A popular recycling method to address this issue is chemical recycling to monomer (CRM) \cite{key3,key4, key5}, which involves breaking down polymers into their original monomer units. This approach is attractive because it enables closed-loop recycling and allows materials to be recovered without a change in properties, as opposed to alternative recycling methods, which can lead to material degradation and loss in performance \cite{key20}. However, many polymers are not readily depolymerizable, making them unsuitable for recycling via CRM. Polymers synthesized via ring-opening polymerization (ROP) are well suited for such sustainable, closed-loop recycling, as their formation from cyclic monomers enables reversible chemistries and efficient depolymerization \cite{key6,key7,key8}. Despite their importance, this class of polymers remains relatively underexplored at the industrial scale. The dataset presented here provides a diverse collection of synthetically accessible and chemically recyclable polymer candidates, informed by deep chemical intuition, thus serving as a valuable resource for a sustainable future.

Virtual Forward Synthesis (VFS) was the first approach employed in this work to generate ROP polymers and has been explored in depth in prior studies \cite{key21, key22}. VFS simulates polymerization by applying established reaction templates to known monomers, thus allowing the generation of hypothetical polymers \cite{key23, key24, key25}. In this work, generalized reaction templates specifically for ROP reactions were compiled from the literature and encoded using SMILES arbitrary target specification (SMARTS) \cite{keysmarts} patterns. Subsequently, these templates were applied to compatible monomers identified by substructure searches of existing databases. Overall, our VFS workflow enabled the generation via enumeration of 8,009,115 hypothetical ROP polymer structures.

The second approach used to generate ROP polymers in this work was the use of language models. Machine learning (ML) has been widely applied in the polymer space to solve the forward problem of predicting polymer properties from their structures. Typically, this involves representing structures numerically using fingerprinting schemes \cite{key9, key10} and training ML models to predict corresponding property values \cite{key11, key12, key13, key14}. In contrast, the reverse problem of generating polymer structures conditioned on desired properties is nascent \cite{VAE}. Recently, language models have emerged as a promising solution and have been widely applied in the molecular domain, leading to the development of models such as SMILES-BERT \cite{key26}, MolBERT \cite{key27}, and SELFIES-TED \cite{key28}, often referred to as molecular foundation models. These models capture the underlying syntax and structure of molecular representations and are capable of both property prediction and molecular generation. Inspired by these advances, generative models for polymers such as polyBART \cite{polybart} and POLYT5 \cite{polyt5} have been recently developed. These have been applied to a range of tasks, including predicting thermal and electronic polymer properties and generating experimentally validated novel polymer designs for dielectric applications. In this work, we applied these models specifically to the generation of ROP polymers, forming the second component of our candidate generation framework. polyBART, based on the BART architecture \cite{key18}, enables property-conditioned polymer generation through latent space exploration, while POLYT5, based on the T5 \cite{key19} architecture, solves this problem via sequence-to-sequence generation. In this work, we built upon these two complementary polymer language models and leveraged them to generate a diverse set of 651,992 hypothetical ROP polymer candidates.

The structures generated in this work using VFS, polyBART, and POLYT5 were subjected to a strict filtering process to ensure the quality of the final dataset. The ability of modern computational workflows to generate millions of hypothetical polymers has far outpaced our ability to assess their chemical feasibility. In practice, experienced chemists rely heavily on implicit intuition on molecular stability, functional group compatibility, and polymerization feasibility to rapidly discard unrealistic candidates. However, this knowledge is rarely formalized and therefore cannot be readily applied at the scale required for high-throughput polymer generation. To bridge this gap, we developed a rule-based chemical heuristics filtering framework that codifies chemical intuition into an automated screening pipeline. The framework applies expert defined criteria to remove structures that are likely to be chemically infeasible, unstable, or incompatible with ring-opening polymerization. By combining ring motif classification with mechanism-aware screening rules, the filter can identify unstable or hazardous functional groups, mechanism-specific interference, and structures lacking plausible ROP ring motifs. In this way, we provide a conservative yet scalable means to refine the large candidate space, improving the synthetic relevance of the resulting polymers while preserving structural diversity.

We developed our dataset with the goal of guiding the discovery of novel, sustainable, and recyclable polymers. The final dataset consists of 1,087,564 ROP polymers, monomers, their structural class, associated reaction mechanism, commercial availability, and synthetic accessibility score (SAscore). These annotations enable users to filter and prioritize candidate polymers according to their specific design requirements and synthesis constraints. The data is available at the polyVERSE \cite{polyverse} repository and is expected to be continuously expanded in future work. We hope that this dataset will serve as a valuable resource for guiding the usage of chemically recyclable ROP polymers for a variety of applications.

\section*{Methods}
\subsection*{Workflow}

\begin{figure}[H]
\centering
\includegraphics[width=\linewidth]{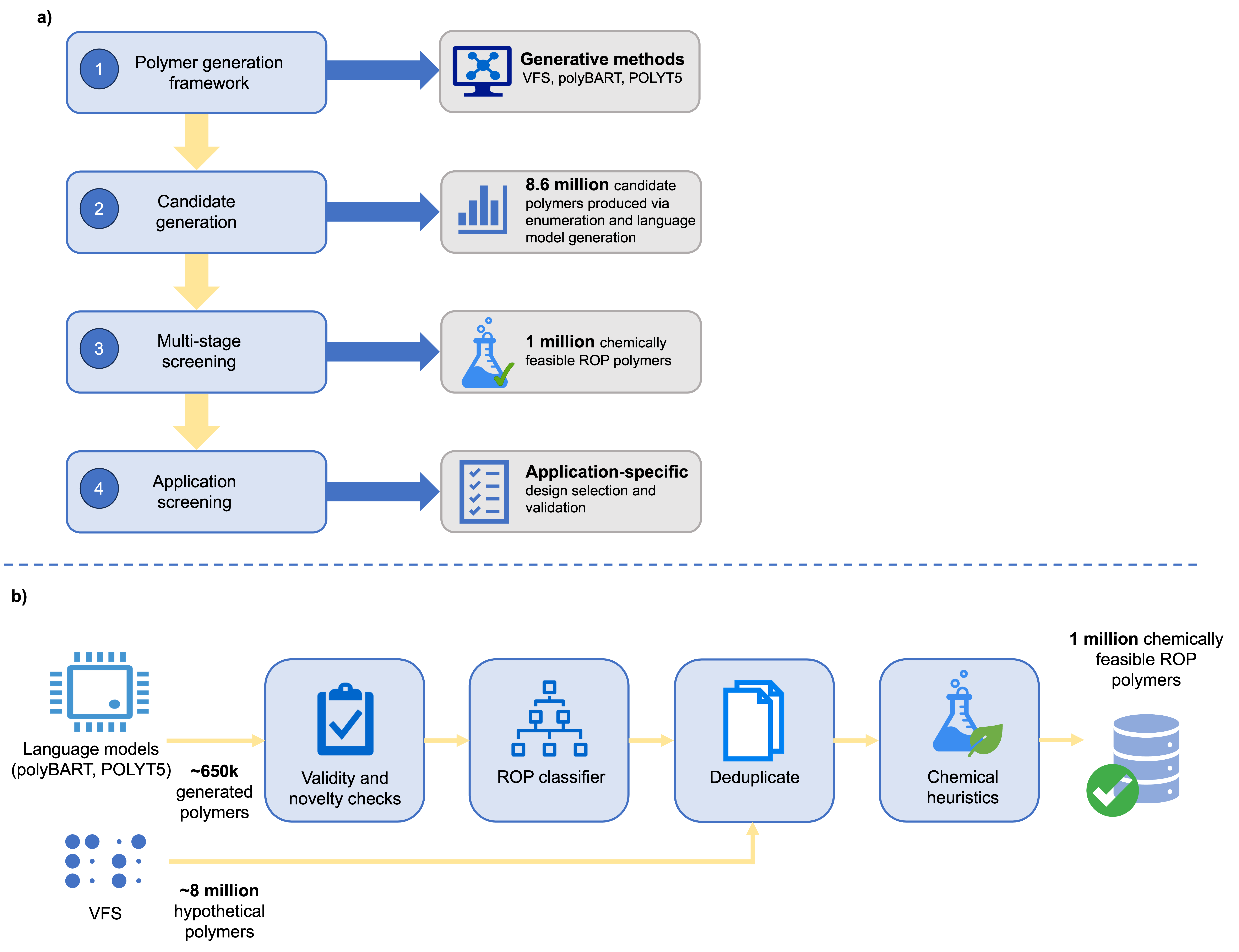}
\caption{\textbf{(a)} Overview of the workflow used to construct the dataset. \textbf{(b)} Schematic of the multi-stage filtering pipeline applied to the resulting candidates. The final dataset comprises 1,454,122 unique ROP polymer SMILES.}
\label{fig:workflow}
\end{figure}

Figure~\ref{fig:workflow}a provides an overview of the overall workflow used to construct the dataset. The pipeline began with the large scale generation of hypothetical ROP polymer candidates using VFS, polyBART, and POLYT5, resulting in millions of candidate structures. The generated candidates were subsequently subjected to a multi-stage validation and filtering procedure to ensure dataset quality, as illustrated in Figure~\ref{fig:workflow}b. First, each candidate was evaluated to confirm both monomer and polymer validity and in the case of the polyBART generated candidates, any overlap with input data was excluded. The language model generations were further screened for compatibility with ROP behavior using ML classification models. The dataset was then deduplicated and finally, a rule-based chemical heuristics filter was applied to remove chemically implausible or unstable structures. Collectively, these filtering steps ensure that the final dataset is composed of chemically valid and feasible structures. To further assess practical synthesizability, the SAscore of each candidate was computed using RDKit \cite{keyRDKit}. 

The final dataset contains 950,019 polymers generated using VFS, 92,217 generated using POLYT5, and 46,457 generated using polyBART.  Additionally, 225 experimentally reported ROP polymers curated from the literature and the PolyInfo \cite{polyinfo} repository were incorporated into the dataset. Tables~\ref{tab:rop_categories} and~\ref{tab:mechanism_profiles} present the distribution of monomer classes and the polymerization reaction mechanisms represented in the data. Our dataset presents 1,087,564 polymer and corresponding monomer structures in the standard Simplified Molecular Input Line Entry System (SMILES) format \cite{keySMILES}, with [*] denoting polymer endpoints. We anticipate that this diverse collection of structures will provide a foundation for future data-driven exploration, screening, and design of ROP polymers across a broad range of applications.

\subsection*{Dataset Construction}
\begin{figure}[H]
\centering
\includegraphics[width=0.8\linewidth]{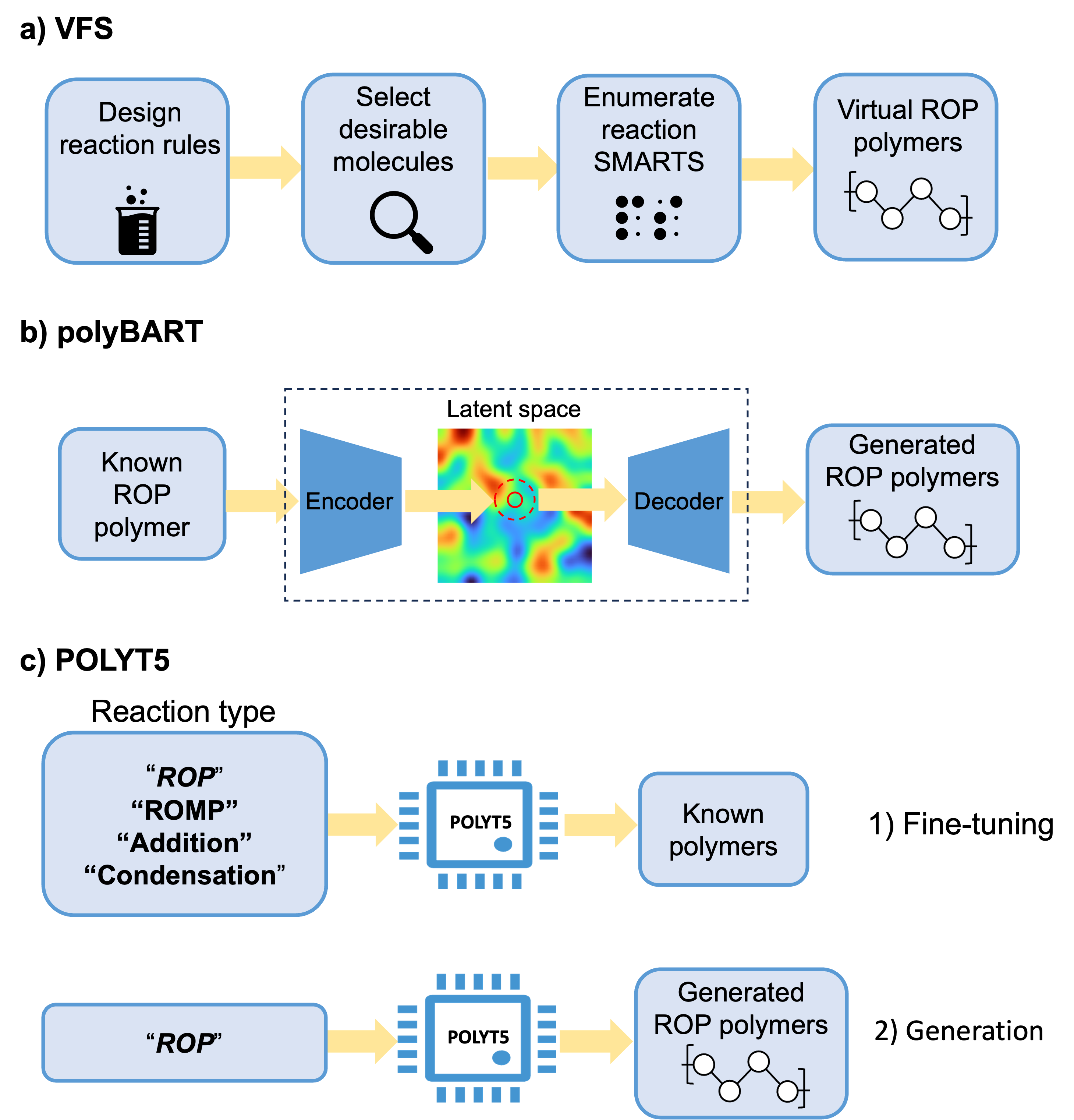}
\caption{Overview of the three approaches used to generate ROP polymers in this work. \textbf{(a)} VFS enumerates polymers from known monomers and predefined reaction templates. \textbf{(b)} polyBART generates novel candidates through latent-space exploration. \textbf{(c)} POLYT5 generates polymers conditioned on the specified reaction class and is prompted with ROP during inference.}
\label{fig:generative_methods}
\end{figure}

As described in detail in the following sections, we used three approaches: VFS, polyBART, and POLYT5, to produce the dataset. A schematic overview of these methods is presented in Figure~\ref{fig:generative_methods}.
\subsection*{Virtual Forward Synthesis}
VFS is a reaction-aware virtual synthesis method for designing polymers from commercial monomer libraries and is executed using the RxnChainer (Reaction Chainer) tool \cite{202605.0977}. This approach has been used extensively in the past to develop databases of synthetically accessible polymers, several of which have been synthesized, and experimentally validated for applications including dielectric energy storage and sustainable food packaging\cite{key_polyuniverse, key_polysulfate, phan2026}. The VFS process involves several key steps. To begin, we designed reaction rules by compiling general templates for polymerization reactions of ring-opening polymers from the literature. These templates involve reacting functional groups and spacer groups with generalized reactions. Based on the established reaction templates, we next performed an RDKit substructure search of \textasciitilde30 million commercially available reactant molecules in databases such as TSCA\cite{keyTSCA}, eMolecules \cite{key_emolecules}, and ChEMBL \cite{key_chembl} to identify those that are compatible with each reaction. After identifying the reactants, we established reaction rules for ring-opening polymerization and encoded them using reaction SMARTS to obtain the corresponding polymer. The implemented SMARTS were sufficiently general to accurately capture the chemical diversity of molecules across different databases, enabling the synthesis of accurate polymers from monomer reactants. Using these SMARTS patterns, we enumerated the SMILES of compatible reactants for each ROP reaction, thereby obtaining hypothetical polymer SMILES for each polymer class.
 
The polymers generated through the VFS workflow are synthetically accessible by construction, as they are derived from known reaction pathways and commercially available or known monomers. Previous applications of this procedure produced a database of \textasciitilde7.4 million hypothetical polymers, which was subsequently released\cite{key21, key_ropdata}. Here, we extended this effort by expanding the database to \textasciitilde8 million hypothetical polymer candidates, which served as the starting search space for the filtering, screening, and down-selection procedures described in this work.

\subsection*{polyBART}

polyBART is a polymer expert foundation model developed by transfer learning from the SELFIES-TED molecular model to the polymer domain. In this work, we applied polyBART specifically to the generation of ROP polymers. The development of polyBART began with the pre-training stage, during which the model learns the underlying language of polymers. We used PSELFIES, a polymer-specific extension of Self-Referencing Embedded Strings (SELFIES) \cite{keySELFIES}, introduced in prior work \cite{polyt5}, to develop polyBART. The pre-training corpus spans a wide range of polymer classes, including polyethers, polyacetals, polyesters, and polyamides, to name a few. These polymers were generated using VFS, by systematically applying rections including ring-opening metathesis polymerization (ROMP) \cite{young2011}, click polymerization, and Sulfur Fluoride Exchange (SuFEx) reactions, to construct a large and chemically diverse collection of synthetically accessible polymers. polyBART was initialized from the SELFIES-TED molecular model, which is available in two variants, small (\textasciitilde2 million parameters), and large (\textasciitilde350 million parameters). Owing to its efficiency and strong performance, we adopted the small model as the starting point for polyBART. We then trained polyBART on our PSELFIES dataset using Masked Language Modeling (MLM) for 3 epochs on \textasciitilde287 million PSELFIES. Our transfer learning approach enabled us to leverage the extensive chemical knowledge encoded in models pre-trained on large molecular datasets, while progressively adapting the model to the polymer domain. 

As described in our previous work \cite{polybart}, new polymers were generated by exploring the latent space learned by polyBART. Practically, this was achieved by adding Gaussian noise to the encoder embedding of an ROP polymer, thus enabling the exploration of nearby regions in the latent space. Noise was applied multiple times to generate a set of perturbed embeddings, which were then decoded using the polyBART decoder. By sampling locally around the embedding of an ROP polymer, polyBART allows for the generation of structurally related candidates and remaining within a region of the latent space associated with desirable characteristics, while simultaneously introducing diversity. The resulting candidates were further subjected to filtering steps to ensure validity and quality of the generations as described in the \textbf{Data Filtering} section.

\subsection*{POLYT5}
POLYT5 is a polymer foundation model trained on \textasciitilde
100 million polymers represented in the PSELFIES format, enabling both property prediction and generative polymer design. To extend its capabilities toward reaction-specific polymer generation, we employed the POLYT5 base model (\textasciitilde7.5 million parameters) and fine-tuned it using a labeled dataset of polymerization reactions. The dataset encompassed four reaction classes: ROP, ROMP, addition, and condensation polymerizations. It included polymers reported in the literature as well as those generated using established reaction templates and known monomers, comprising 2,354,364 ROP, 127,496 ROMP, 6,491 condensation, and 2,088 addition examples. Although our primary objective was to generate ROP polymers, we intentionally incorporated polymers from other reaction classes to provide broader chemical context and enable the model to better distinguish between polymerization mechanisms. During fine-tuning, the input consisted of reaction-type keywords, and the output was the corresponding polymer structure in SELFIES representation. The training protocol followed the same sequence-to-sequence framework as described in our prior work.\cite{polyt5} For ROP polymer generation, the fine-tuned POLYT5 model was prompted with the desired reaction type and sampled under varying decoding conditions, including temperature, which controls generation randomness, nucleus sampling (top-p), which limits sampling to the most probable tokens, and the number of fine-tuning epochs.

\subsection*{Generative Capabilities}

\begin{figure}[H]
\centering
\includegraphics[width=\linewidth]{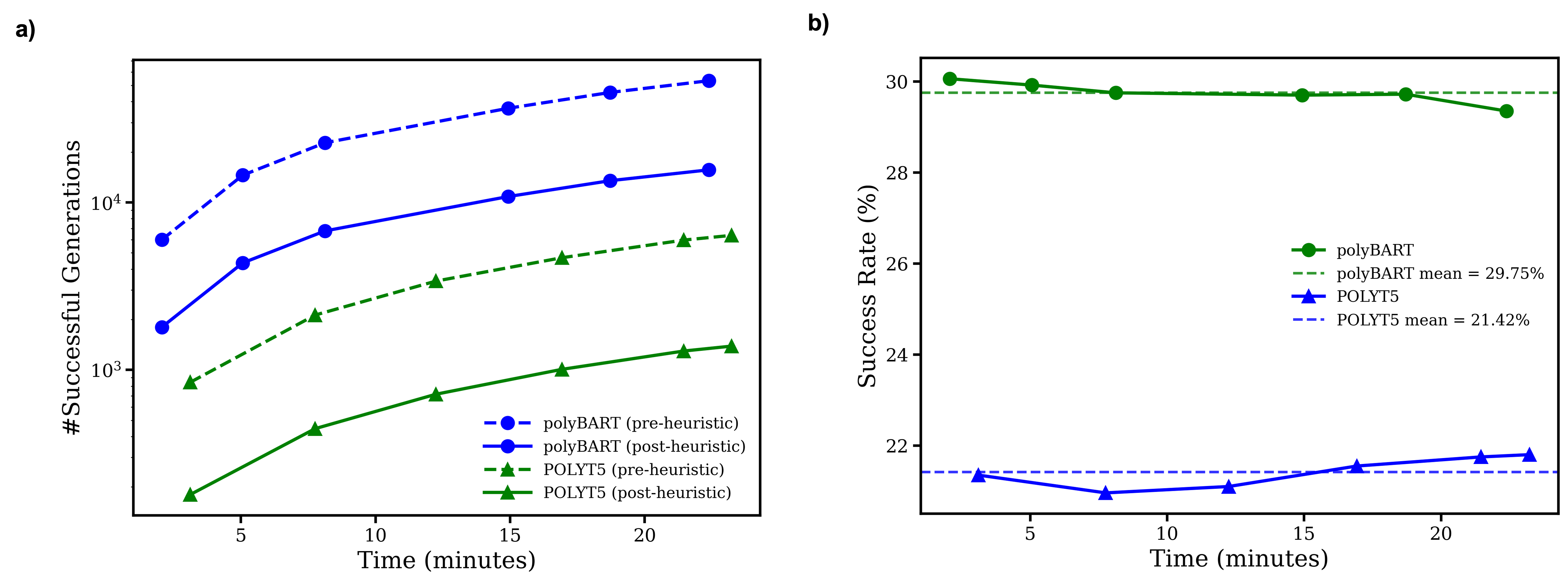}
\caption{Comparison of the generative performance of polyBART and POLYT5. \textbf{(a)} Cumulative number of successful generations as a function of time before and after heuristic filtering. The y-axis is displayed on a logarithmic scale. \textbf{(b)} Success rate of generated candidates, defined as the fraction of structures that survive heuristic filtering. Dashed lines indicate the mean success rate for each model.}
\label{fig:comparison}
\end{figure}

We compared the performance of the language models in terms of generation efficiency and the effect of applying heuristic filters as depicted in Figure~\ref{fig:comparison}. Performance was evaluated using two metrics: (i) the cumulative number of successful generations as a function of time and (ii) the success rate. Successful generations are defined as candidates that pass the initial validity and screening checks in our pipeline and success rate is the percentage of generations that survive the heuristic filter.

The results show that polyBART generates more valid candidates over time compared to POLYT5, both before and after filtering. This can be attributed to a number of reasons. First, polyBART supports batched latent-space sampling and decoding, allowing many candidates to be generated simultaneously. POLYT5 generation in our implementation is performed sequentially, with each candidate generated and evaluated individually. Second, polyBART has a smaller model size (\textasciitilde2 million parameters) compared to POLYT5 (\textasciitilde7.5 million parameters), leading to faster inference. Together, these factors enable polyBART to explore a significantly larger number of candidate structures per unit time, leading to higher overall throughput. PolyBART also exhibits a higher success rate, with an average of 29.75\%, compared to 21.42\% for POLYT5.

\subsection*{Data Filtering}

\subsection*{Validity and Novelty}
The generated structures underwent a systematic filtering process, as depicted in Figure~\ref{fig:workflow}b. The first step in the filtering pipeline checks whether the structures generated by polyBART and POLYT5 are both valid molecules and valid polymers. This step is essential, as language models are inherently stochastic and not explicitly guided by chemical rules, and therefore can at times generate invalid or chemically implausible structures. Molecular validity was assessed by converting the generated SELFIES strings to SMILES using the SELFIES decoder and then evaluating them using RDKit. Polymer validity was verified by checking the endpoint characteristics. This involves ensuring that each structure contains exactly two endpoints, each with valency one, and that both endpoints have consistent bond types. For polyBART, which generates candidates from a seed ROP polymer, we additionally enforced novelty by removing generated structures identical to the input seeds. The generated structures were canonicalized using the \textit{canonicalize\_psmiles} package \cite{kuenneth2024} to reduce multiple representations of the same polymer, after which duplicate generations were removed. This filter ensures that the retained candidates are chemically valid, novel, and suitable for further evaluation.

\subsection*{ROP Classifier}
To further refine the generated candidates, we passed them through an ROP classifier. This classifier is an ML model that takes a polymer structure as input and predicts its corresponding reaction class. Given the stochastic nature of language model generation, candidates that pass initial validity and novelty checks may not necessarily correspond to true ROP polymers. The ML classifiers, therefore, serve as an additional filtering step, increasing confidence that the retained candidates are consistently ROP. This step can be skipped for polymers generated via VFS, since VFS explicitly constructs polymers through predefined ROP reaction rules and commercially available monomers, resulting in candidates that are already highly likely to be consistent with ROP chemistries.

In the case of polyBART, the classifier was implemented as a neural network that operates on its encoder embeddings of the polymer and outputs the predicted reaction type as a one-hot encoded vector. The classifier was trained on a curated dataset of 2,000 polymers paired with their corresponding polymerization reaction classes collected from the literature, comprising 500 examples each of Addition, Condensation, ROP, and ROMP. For POLYT5, the reaction classifier was implemented as a sequence-to-sequence POLYT5 model trained on this dataset, but with the input–output mapping reversed. Specifically, the model takes a polymer SELFIES string as input and predicts the corresponding reaction class. The polyBART classifier achieves an overall test accuracy of 89\%, while POLYT5 achieves 96\%. Notably, the polyBART classifier attains 88\% test accuracy on the ROP class, with POLYT5 achieving 91\%. These results demonstrate that both models are capable of reliably distinguishing between polymerization mechanisms and validate their use for the filtering of ROP generations.

\subsection*{Chemical Heuristics}
%Ask Wei to add references
At the scale considered in this work, manual inspection of the generated candidates is impractical. The size of the candidate pool, coupled with the novelty of the generations, makes it infeasible for a chemist to individually evaluate the 8.6 million generations for consistency with ROP chemistries. The chemical heuristics rules developed here are designed to automate this screening process by encoding established chemical intuition through SMARTS-based pattern matching. These rules enable the rapid identification and removal of candidates that are unlikely to be compatible with ROP mechanisms. In this way, we are able to benefit from large scale exploration of the ROP design space while maintaining a high degree of confidence in the retained structures.

\begin{figure}[H]
\centering
\includegraphics[width=\linewidth]{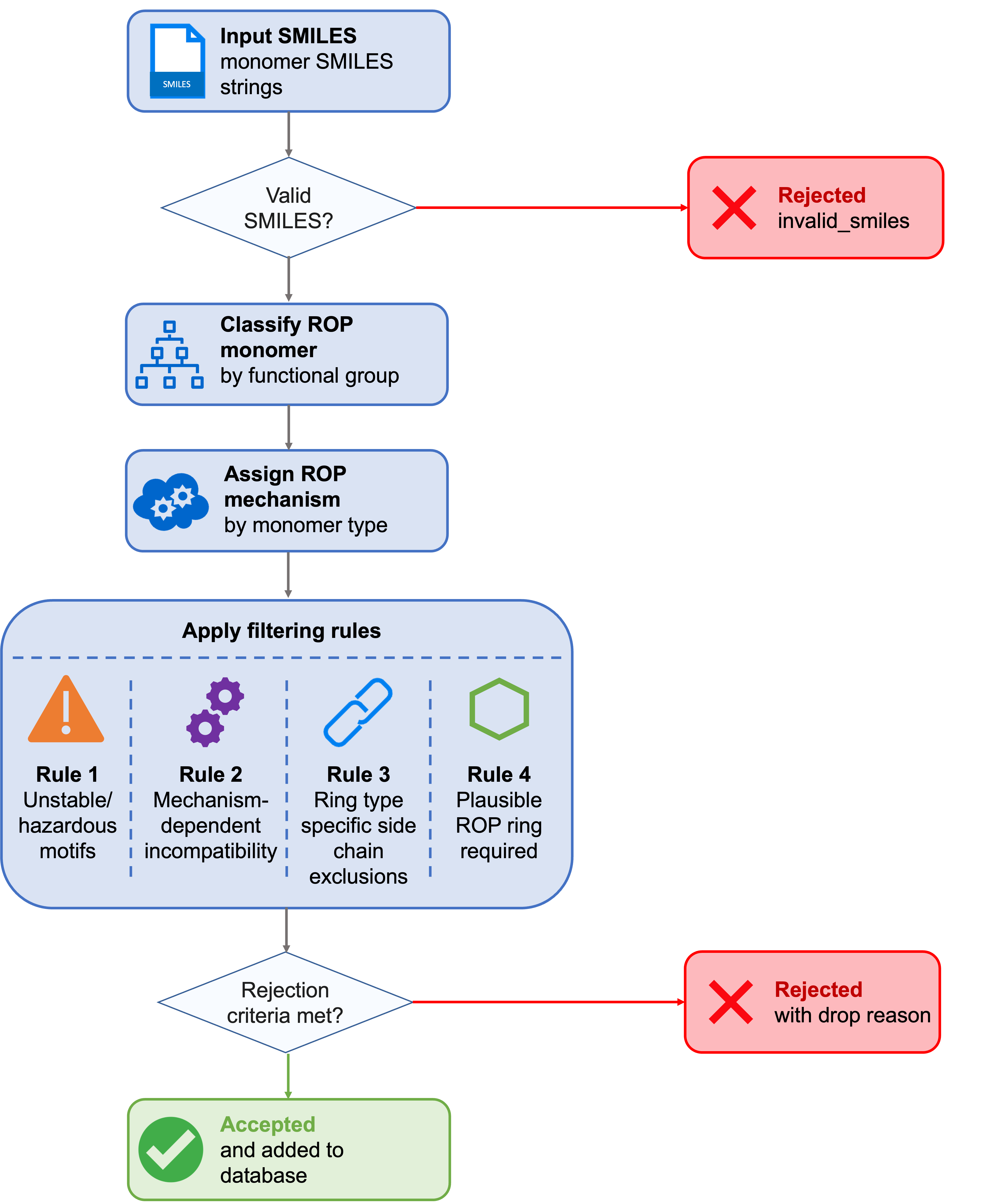}
\caption{Overview of the chemical heuristics filtering workflow. Polymer candidates are validated, assigned a monomer class and polymerization mechanism, and subsequently screened using the four chemistry-informed rules.}
\label{fig:heuristics}
\end{figure}

This filter consists of the following major components: (1) ring motif classification and ROP-scope checking; (2) removal of molecules containing unstable or hazardous groups; (3) removal of molecules containing mechanism-dependent or type-specific interfering functional groups; and (4) removal of molecules with no plausible ROP ring motif or low-strain/non-ROP ring structures. These rules are intentionally conservative and are designed to remove structures that are likely to fail for clear chemical reasons before databasing and downstream application-based screening. They are meant for high-throughput candidate space cleanup rather than for making absolute claims about whether a given molecule can ever be polymerized under specialized conditions. An overview of this filter and its four rules is shown in Figure ~\ref{fig:heuristics}. The SMARTS patterns used to identify structures for removal by this filter are listed in Tables~\ref{tab:rule1}--\ref{tab:rule4}.

The preliminary step in this workflow was to construct monomer SMILES for each polymer by cyclizing the repeat unit through the connection of the terminal [*]. Before applying any exclusion rules, each input monomer SMILES was first parsed with RDKit and inspected for ring motifs. They were then assigned to a putative ROP type when they match a priority motif or ring-bond pattern. The current motif library covers \textit{N}-carboxyanhydrides (NCAs), \textit{N}-thiocarboxyanhydrides (NTAs), \textit{O}-carboxyanhydrides (OCAs), cyclic anhydrides, cyclic carbonates, thiolactones, lactams, lactones/lactides, 2-oxazolines, aziridines/azetidines, 1,2-dithiolanes, cyclic phosphoesters, cyclic siloxanes, cyclic silazanes, cyclic ketene acetals, vinyl cyclopropane/butane motifs, epoxides, cyclic sulfides, cyclic ethers, and cycloolefins. 

Rule 1 removes molecules containing chemically unstable or potentially hazardous groups. Typical examples include peroxides, ozonides, diazo groups, azides, nitrate esters, nitramines, hydrazones, imines, oximes, sulfonyl azides, acyl azides, azidoformates, perchlorate esters, acyl halides, sulfonyl chlorides, phosphoryl chlorides, N--O, N--S, N=S, N--halogen, X--O, and S--halogen linkages. The main reason for this rule is that these groups may decompose during synthesis, purification, storage, or polymerization. Some of them may also cause safety problems. Therefore, this rule mainly works as a stability and safety filter.

Rule 2 removes molecules that contain functional groups likely to interfere with the expected polymerization mechanism. The filter first assigns a default mechanism profile from the detected ROP type. For example, lactones and cyclic carbonates are checked against anionic, coordination-insertion, and organocatalytic profiles; 2-oxazolines are checked against cationic profiles; cyclic ketene acetals and vinyl cyclopropane/butane motifs are checked against radical profiles; and cycloolefins are assigned to a ROMP profile and are excluded from the dataset. For anionic and organocatalytic profiles, the filter checks groups such as alcohols/phenols, carboxylic acids, thiols, amines, amide N-H, sulfonamide N-H, terminal alkynes, and strongly basic groups. For cationic profiles, it checks Lewis-basic or nucleophilic groups such as amines, aza-aromatics, imidazole-like motifs, thiols, alcohols/phenols, and phosphine/amine Lewis bases. For coordination-insertion profiles, it checks groups that may bind or poison metal centers, including amines, aza-aromatics, imidazole-like motifs, phosphines, thiols, thioethers, carboxylates, carboxylic acids, and beta-dicarbonyl-like chelating motifs. The reason for this rule is that reactive or strongly coordinating functional groups can interact with the catalyst, initiator, propagating species, or reactive intermediates, leading to catalyst poisoning, initiation/propagation interference, chain-transfer-like behavior, or competing side reactions. This rule is therefore mechanism-aware rather than a single global side-chain blacklist.

Rule 3 removes molecules when their non-ring side chains contain functional groups that are problematic for the assigned monomer type. The current side-chain patterns include esters, carbonates, thioesters, amides, disulfides, alkenes, alkynes, isocyanates, and aldehydes. These groups are not applied globally, instead each monomer type has its own exclusion profile. For example, lactones, lactams, cyclic carbonates, epoxides, aziridines, and cyclic ethers are checked against several acyl/carbonate/thioester/disulfide/isocyanate-type side-chain motifs, whereas cyclic ketene acetals and vinyl cyclopropane/butane motifs are mainly checked for side-chain disulfides, alkenes, and alkynes. Cycloolefins currently have no type-specific side-chain exclusion in this rule. The side-chain match is counted only when the entire matched motif is outside the ring system. This design prevents the filter from incorrectly removing molecules because the intended ROP-active ring itself contains an ester, carbonate, amide, thioester, or related functional group. 

Rule 4 removes molecules that are outside the intended ROP chemical space. Molecules with no plausible ROP ring motif, or only cycloalkane/cycloketone-like rings are removed. The rule also removes specific low-strain motifs, including tetrahydropyran-like cyclic ethers and gamma-thiolactone-like structures, when these motifs are expected to have insufficient ring-opening driving force under general ROP conditions. The reason is that ring strain release and/or favorable ring-opening thermodynamics are important driving forces for many ROP systems. If a molecule lacks a suitable ring motif, contains only an unreactive saturated carbocycle, or has a low-strain ring, ring opening may be weak, reversible, or unfavorable. Such monomers may therefore show low conversion or poor polymerization behavior under general screening conditions. This rule focuses on the ring-opening driving force and ROP-scope definition, rather than on side reactivity caused by substituents.

\subsection*{Application-specific filtering}

\begin{figure}[H]
\centering
\includegraphics[width=\linewidth]{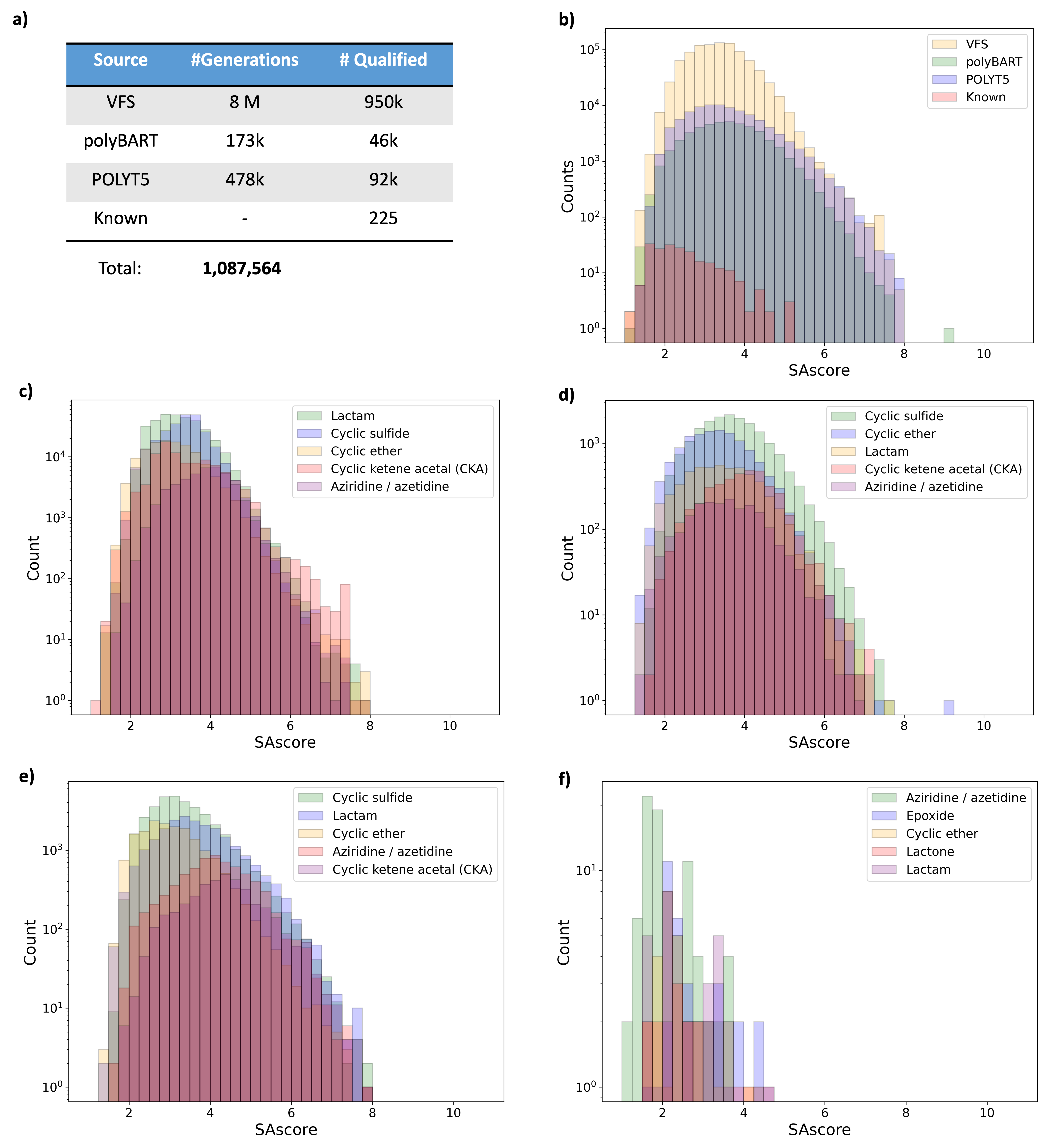}
\caption{\textbf{(a)} Total number of generations produced by each source and number of polymers retained in the final dataset. \textbf{(b)} SAscore distributions grouped by generation source. \textbf{(c--f)} SAscore distributions for the five most prevalent monomer classes within the VFS, polyBART, POLYT5, and known polymer subsets, respectively. The y-axis is displayed on a logarithmic scale.
}
\label{fig:sascores}
\end{figure}

We finally predicted the synthetic accessibility of the surviving ROP polymers using the RDKit implementation of the SAscore \cite{keySA}. This metric estimates the ease of synthesis of a molecule by balancing the prevalence of constituent molecular fragments in known compounds against structural features such as molecular size, stereochemistry, and ring complexity. The resulting SAscore ranges from 1 (highly accessible) to 10 (highly challenging to synthesize). Since the RDKit implementation is defined for molecular rather than polymer representations, each polymer repeat unit was first capped with hydrogen atoms prior to SAscore calculation. We observed an average SAscore of 3.3 across the final generated structures, with the full distribution shown in Figure~\ref{fig:sascores}. For comparison, the commercially available polymers in our dataset exhibit an average SAscore of 2.5, which is consistent with the average SAscore of approximately 3 for known molecules. The close agreement between the distributions of the generations and the known polymers suggests that the majority of retained candidates possess synthetic accessibility comparable to that of experimentally realizable molecules, supporting the practical feasibility of the generated polymer designs.

Given the dataset of polymers, structural classes, mechanism profiles, and SAscores, chemists can now perform application-specific screening to identify candidates of interest for further analysis and potentially laboratory synthesis and validation. Since the primary objective of this work is to present and describe the dataset itself, we do not pursue such application-focused studies here. However, we suggest a few promising application domains for this framework and dataset, which will be explored in detail in future work. The first is the identification of recyclable alternatives to widely used commodity and packaging plastics, such as polyethylene (PE), polystyrene (PS), and polyethylene terephthalate (PET), whose widespread use and resistance to degradation have made them major contributors to long-term environmental pollution and landfill accumulation \cite{keymicroplastics, keylandfill, keygeyer, keymicroplastics2}. A second, less obvious application is the development of high performance polymer matrices for fiber-reinforced composites used in automotive and aerospace sectors. Existing composites are challenging to recycle and are often downcycled through energy recovery processes, resulting in the loss of high-value fibers \cite{keyCompositeRecycle, keyCompositeRecycleAviation}. Recyclable ROP alternatives can enable the development of polymer composite matrices specifically designed for end-of-life deconstruction and material recovery. Finally, we consider additive manufacturing, an area where there is growing demand for recyclable feedstock materials \cite{key3DPrintingRecycle, key3DPrintingPhotopolymer}. Here, ROP polymers have huge potential to provide a promising route toward high performance printing materials that can be repeatedly recovered, reprocessed, and reused. Overall, we hope that the framework and dataset presented in this work will contribute to the long-term goal of establishing a truly circular plastics economy.

\section*{Data Record}
The complete dataset of 1,087,564 polymer entries can be accessed in Parquet format through the polyVERSE repository at https://github.com/Ramprasad-Group/polyVERSE/tree/main/ROP. Each row of the dataset contains the following information: 

\texttt{smiles}: the polymer SMILES representation

\texttt{canonical\_smiles}: the canonicalized polymer SMILES 

\texttt{source}: the method by which the structure was generated or curated

\texttt{monomer\_smiles}: the corresponding cyclic ROP monomer SMILES

\texttt{rop\_type}: the monomer class assigned by the chemical heuristics filter 

% \texttt{rop\_motif\_reason}: the rationale for the assigned ROP class and the detected structural motif

\texttt{mechanism\_profile}: the polymerization mechanism profile predicted by the chemical heuristics filter

\texttt{commercial\_availability}: a Boolean indicator of whether the corresponding monomer is commercially available

\texttt{SAscore}: the RDKit synthetic accessibility score of the hydrogen-capped polymer SMILES

\section*{Technical Validation}

The filtering procedures described in the preceding sections ensure that the final dataset consists of valid, unique, and chemically feasible ROP structures. As an additional validation step, a subset of the generated structures was reviewed by an expert to assess chemical feasibility and identify promising candidates. However, given the scale of the dataset, comprehensive manual evaluation of every structure is impractical. Therefore, we instead validated our approach by comparing the generations against databases of known and commercially available ROP monomers. Overlap with experimentally reported monomers provides evidence that the framework successfully captures relevant regions of the ROP design space while retaining the ability to discover novel candidates beyond existing databases. Comparing the generated monomers against the TSCA inventory of commercially available compounds revealed 438 matches covering a broad range of monomer classes including cyclic ethers, latcones, and lactams to name a few. Of these, 403 were generated by VFS, 59 by polyBART, and 13 by POLYT5, with several monomers independently identified by multiple approaches.

\begin{figure}[H]
\centering
\includegraphics[width=\linewidth]{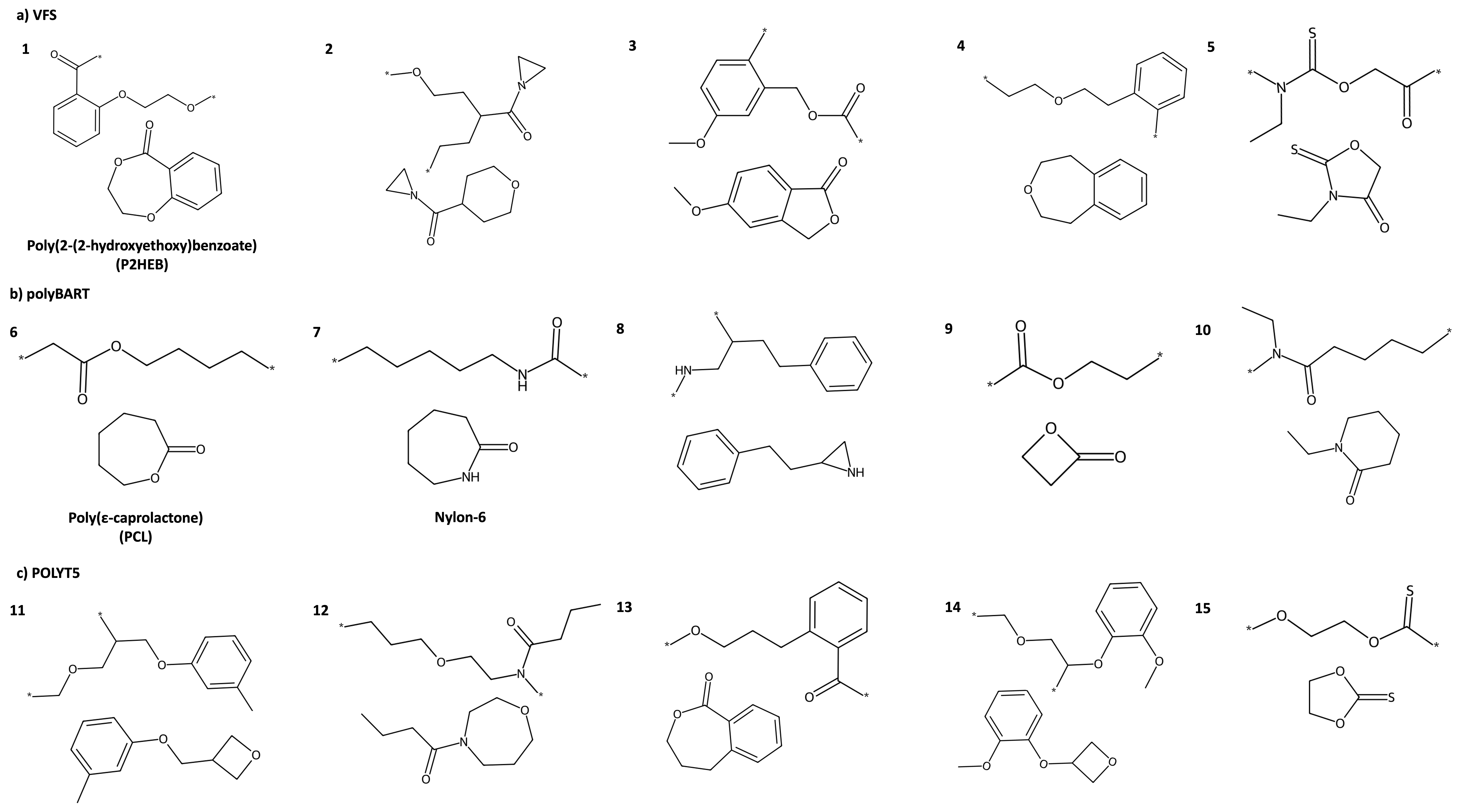}
\caption{Polymer repeat unit structures (top) and corresponding monomer structures (bottom) for selected promising candidates generated using \textbf{(a)} VFS, \textbf{(b)} polyBART, and \textbf{(c)} POLYT5. P2HEB, PCL, and Nylon-6 are rediscovered by the generation framework, providing retrospective validation of its ability to recover known, experimentally realized ROP polymers while also proposing novel candidates.}
\label{fig:examples}
\end{figure}

Many of the commercially available monomers generated by our framework have been widely reported and studied experimentally. For example, polyBART generated poly($\varepsilon$-caprolactone) (PCL), a well established biodegradable polyester produced via the ring-opening polymerization of $\varepsilon$-caprolactone \cite{keyPCLReview}. PCL is an extensively studied ROP polymer owing to its favorable mechanical properties, miscibility with other polymer systems, and excellent biodegradability, characteristics that have enabled its widespread use in biomedical applications \cite{keyPCLBiomedical}. Nylon-6 is another well-established ROP polymer generated by polyBART. It is produced through the ring-opening polymerization of $\varepsilon$-caprolactam and is one of the most widely used engineering thermoplastics \cite{keyNylon6ROP}. Poly(2-(2-hydroxyethoxy)benzoate) (P2HEB) \cite{macdonald2016}, identified by VFS and polyBART, is another experimentally established ROP polymer present in the dataset. P2HEB is notable for its remarkable chemical recyclability, undergoing highly selective depolymerization to its parent monomer. Representative examples of promising ROP candidates identified within the dataset are shown in Figure~\ref{fig:examples}. Collectively, these examples highlight the ability of our framework to rediscover known and experimentally validated ROP systems, providing confidence in the quality of the generated dataset and its potential to guide the discovery of novel sustainable polymers.

\section*{Data Availability}

The complete ROP polymer dataset generated in this work is publicly available at the polyVERSE repository in Parquet format. The dataset includes the polymer and corresponding monomer SMILES representations together with their associated annotations.

\section*{Code Availability}
The code produced during the current study is not publicly available due to IP protection being considered at the authors' institution. 

\section*{Acknowledgements}
This work was supported by the National Science Foundation through grant 2515411 and by the Office of Naval Research  through grant N00014-19-1-2586.
 
\section*{Author Contributions}
A.S. developed the polyBART model, generated the corresponding polymer candidates, filtered and compiled the overall dataset, prepared the figures and tables, and wrote the manuscript. W.X. developed the chemical heuristics filter and contributed to the figures, tables, and writing. H.S. developed the POLYT5 model, generated the corresponding polymer candidates, and contributed to writing the manuscript. S.S.S. extended the VFS polymer candidate space and contributed to writing the manuscript. W.G. provided guidance on developing the chemical heuristics filter. R.R. conceived the study, supervised the research, and oversaw the preparation and final approval of the manuscript.

\section*{Competing Interests}
The authors declare no competing interests.

\newpage

\newpage

\begin{table}[H]
\centering
\begin{tabular}{lr}
\hline
\textbf{Monomer Class} & \textbf{Count} \\
\hline
Lactam & 357,366 \\
Cyclic sulfide & 307,905 \\
Cyclic ether & 155,805 \\
Cyclic ketene acetal (CKA) & 104,181 \\
Aziridine / azetidine & 53,311 \\
Cyclic carbonate & 45,969 \\
Lactone / lactide (cyclic ester) & 30,198 \\
Epoxide / small cyclic ether & 10,997 \\
Thiolactone & 7,484 \\
Cyclic anhydride & 6,452 \\
Vinyl-cyclopropane/butane motif & 5,027 \\
1,2-Dithiolane & 1,729 \\
2-Oxazoline & 726 \\
Cyclic phosphoester & 350 \\
Cyclic silazane & 46 \\
Cyclic siloxane & 16 \\
N-Thiocarboxyanhydride (NTA) & 2 \\
\hline
\end{tabular}
\caption{Distribution of monomer classes in the dataset, categorized according to functional group.}
\label{tab:rop_categories}
\end{table}

\begin{table}[H]
\centering
\begin{tabular}{lr}
\hline
\textbf{ROP Reaction Mechanism} & \textbf{Count} \\
\hline
Cationic polymerization & 886,172 \\
Anionic polymerization & 821,825 \\
Organocatalytic polymerization & 449,550 \\
% ROMP (ring-opening metathesis polymerization) & 366,558 \\
Radical ring-opening polymerization & 109,208 \\
Coordination-insertion polymerization & 82,969 \\
\hline
\end{tabular}
\caption{Distribution of predicted ROP reaction mechanisms in the dataset. Mechanism assignments were determined based on monomer class and known ROP reactivity patterns.}
\label{tab:mechanism_profiles}
\end{table}

\begin{table}[H]
\centering
\small
\begin{tabular}{lll}
\hline
\textbf{Category} & \textbf{Excluded motif} & \textbf{SMARTS pattern} \\
\hline
Unstable / highly labile motifs & imine C=N & \texttt{[CX3;!\$([CX3](=O)N)]=[NX2,NX3;!\$([N-])]} \\
Unstable / highly labile motifs & oxime & \texttt{[CX3]=N-[OX1H]} \\
Unstable / highly labile motifs & hydrazone & \texttt{[CX3]=N-[NX3]} \\
Unstable / highly labile motifs & N--O single bond & \texttt{[N]-[O]} \\
Unstable / highly labile motifs & N--S single bond & \texttt{[N]-[S]} \\
Unstable / highly labile motifs & N=S double bond & \texttt{[N]=[S]} \\
Unstable / highly labile motifs & acyl halide & \texttt{[CX3](=O)[F,Cl,Br,I]} \\
Unstable / highly labile motifs & sulfonyl chloride & \texttt{[SX4](=O)(=O)Cl} \\
Unstable / highly labile motifs & phosphoryl chloride & \texttt{[PX4](=O)(Cl)} \\
Unstable / highly labile motifs & azidoformate & \texttt{[CX3](=O)O-N=[N+]=[N-]} \\
Hazardous / high-energy motifs & peroxide & \texttt{[O;X2]-[O;X2]} \\
Hazardous / high-energy motifs & ozonide & \texttt{[O;X2]-[O;X2]-[O;X2]} \\
Hazardous / high-energy motifs & diazo & \texttt{[CX3]=[N+]=[N-]} \\
Hazardous / high-energy motifs & azide & \texttt{N=[N+]=[N-]} \\
Hazardous / high-energy motifs & nitrate ester & \texttt{[OX2]-[NX3](=O)=O} \\
Hazardous / high-energy motifs & nitramine & \texttt{[NX3]-[NX3](=O)=O} \\
Hazardous / high-energy motifs & sulfonyl azide & \texttt{[SX4](=O)(=O)-N=[N+]=[N-]} \\
Hazardous / high-energy motifs & acyl azide & \texttt{[CX3](=O)-N=[N+]=[N-]} \\
Hazardous / high-energy motifs & perchlorate ester & \texttt{[OX2]-[Cl](=O)(=O)(=O)-[OX2]} \\
Hazardous / high-energy motifs & N--X bond & \texttt{[N;!\$([N-])]-[F,Cl,Br,I]} \\
Hazardous / high-energy motifs & X--O bond & \texttt{[F,Cl,Br,I]-[OX2]} \\
Hazardous / high-energy motifs & S--X bond & \texttt{[S]-[F,Cl,Br,I]} \\
\hline
\end{tabular}
\caption{SMARTS patterns used in Rule 1 to exclude molecules containing chemically unstable or potentially hazardous functional groups.}
\label{tab:rule1}
\end{table}

\begin{table}[H]
\centering
\small
\begin{tabular}{lll}
\hline
\textbf{ROP mechanism} & \textbf{Excluded motif} & \textbf{SMARTS pattern} \\
\hline
Anionic ROP & alcohol / phenol & \texttt{[OX2H]} \\
Anionic ROP & carboxylic acid & \texttt{[CX3](=O)[OX2H]} \\
Anionic ROP & thiol & \texttt{[SX2H]} \\

Anionic ROP & primary / secondary amine &
\begin{tabular}[t]{@{}l@{}}
\texttt{[NX3;H1,H2;!\$([NX3][CX3]=O);} \\
\texttt{!\$([NX3][SX4](=O)=O);!R]}
\end{tabular} \\

Anionic ROP & amide N--H &
\begin{tabular}[t]{@{}l@{}}
\texttt{[NX3;H1,H2;!R]}\\
\texttt{[CX3](=O)}
\end{tabular} \\

Anionic ROP & sulfonamide N--H &
\begin{tabular}[t]{@{}l@{}}
\texttt{[NX3;H1,H2;!R]}\\
\texttt{[SX4](=O)(=O)}
\end{tabular} \\

Anionic ROP & terminal alkyne & \texttt{[CX2H]\#[CX2]} \\

\hline

Cationic ROP & amine &
\begin{tabular}[t]{@{}l@{}}
\texttt{[NX3;!\$([NX3][CX3]=O);} \\
\texttt{!\$([NX3][SX4](=O)=O);!R]}
\end{tabular} \\

Cationic ROP & pyridine / aza-aromatic & \texttt{[nH0]} \\
Cationic ROP & imidazole-like group & \texttt{[nH0]1[c,n][c,n][c,n]1} \\
Cationic ROP & thiol & \texttt{[SX2H]} \\
Cationic ROP & alcohol / phenol & \texttt{[OX2H]} \\
Cationic ROP & strong Lewis base & \texttt{[PX3,NX3;!R]} \\

\hline

Coordination-insertion ROP & amine &
\begin{tabular}[t]{@{}l@{}}
\texttt{[NX3;!\$([NX3][CX3]=O);} \\
\texttt{!\$([NX3][SX4](=O)=O);!R]}
\end{tabular} \\

Coordination-insertion ROP & pyridine / aza-aromatic & \texttt{[nH0]} \\
Coordination-insertion ROP & imidazole-like group & \texttt{[nH0]1[c,n][c,n][c,n]1} \\
Coordination-insertion ROP & phosphine & \texttt{[PX3]} \\
Coordination-insertion ROP & thiol & \texttt{[SX2H]} \\
Coordination-insertion ROP & thioether & \texttt{[SX2;H0;!R]} \\
Coordination-insertion ROP & carboxylate & \texttt{[CX3](=O)[O-]} \\
Coordination-insertion ROP & carboxylic acid & \texttt{[CX3](=O)[OX2H]} \\

Coordination-insertion ROP & chelating $\beta$-dicarbonyl-like group &
\begin{tabular}[t]{@{}l@{}}
\texttt{[CX3](=O)[\#6]}\\
\texttt{[CX3](=O)}
\end{tabular} \\

\hline

Organocatalytic ROP & alcohol / phenol & \texttt{[OX2H]} \\

Organocatalytic ROP & amine &
\begin{tabular}[t]{@{}l@{}}
\texttt{[NX3;!\$([NX3][CX3]=O);} \\
\texttt{!\$([NX3][SX4](=O)=O);!R]}
\end{tabular} \\

Organocatalytic ROP & carboxylic acid & \texttt{[CX3](=O)[OX2H]} \\
Organocatalytic ROP & thiol & \texttt{[SX2H]} \\
Organocatalytic ROP & phenol & \texttt{[OX2H][c]} \\
Organocatalytic ROP & basic group & \texttt{[nH0]} \\

\hline
\end{tabular}
\caption{SMARTS patterns used in Rule 2 to exclude molecules containing functional groups that may interfere with the assigned ROP reaction mechanism.}
\label{tab:rule2}
\end{table}

\begin{table}[H]
\centering
\small
\begin{tabular}{p{2.7cm}p{4.2cm}p{7.0cm}}
\hline
\textbf{ROP monomer type} & \textbf{Excluded side-chain motifs} & \textbf{Corresponding SMARTS patterns} \\
\hline
Lactam & ester; carbonate; thioester; amide; disulfide; isocyanate &
\texttt{ester: [CX3](=O)O[!\$(*=O)]; carbonate: [CX3](=O)O[OX2]; thioester: [CX3](=O)S; amide: [CX3](=O)N; disulfide: S-S; isocyanate: N=C=O} \\
Lactone & ester; carbonate; thioester; disulfide; isocyanate &
\texttt{ester: [CX3](=O)O[!\$(*=O)]; carbonate: [CX3](=O)O[OX2]; thioester: [CX3](=O)S; disulfide: S-S; isocyanate: N=C=O} \\
Thiolactone & carbonate; thioester; disulfide; isocyanate &
\texttt{carbonate: [CX3](=O)O[OX2]; thioester: [CX3](=O)S; disulfide: S-S; isocyanate: N=C=O} \\
Cyclic ether & ester; carbonate; thioester; disulfide; isocyanate &
\texttt{ester: [CX3](=O)O[!\$(*=O)]; carbonate: [CX3](=O)O[OX2]; thioester: [CX3](=O)S; disulfide: S-S; isocyanate: N=C=O} \\
Cyclic carbonate & ester; carbonate; thioester; disulfide; isocyanate &
\texttt{ester: [CX3](=O)O[!\$(*=O)]; carbonate: [CX3](=O)O[OX2]; thioester: [CX3](=O)S; disulfide: S-S; isocyanate: N=C=O} \\
Cyclic ketene acetals / CKA & disulfide; alkene; alkyne &
\texttt{disulfide: S-S; alkene: [C]=[C]; alkyne: [C]\#[C]} \\
NCA & thioester; isocyanate &
\texttt{thioester: [CX3](=O)S; isocyanate: N=C=O} \\
OCA & thioester; isocyanate &
\texttt{thioester: [CX3](=O)S; isocyanate: N=C=O} \\
2-oxazolines & thioester; disulfide; isocyanate &
\texttt{thioester: [CX3](=O)S; disulfide: S-S; isocyanate: N=C=O} \\
Cyclic sulfides & thioester; disulfide; isocyanate &
\texttt{thioester: [CX3](=O)S; disulfide: S-S; isocyanate: N=C=O} \\
1,2-dithiolanes & thioester; disulfide; isocyanate &
\texttt{thioester: [CX3](=O)S; disulfide: S-S; isocyanate: N=C=O} \\
Aziridines & ester; carbonate; thioester; disulfide; isocyanate &
\texttt{ester: [CX3](=O)O[!\$(*=O)]; carbonate: [CX3](=O)O[OX2]; thioester: [CX3](=O)S; disulfide: S-S; isocyanate: N=C=O} \\
Cyclic phosphoesters & thioester; disulfide; isocyanate &
\texttt{thioester: [CX3](=O)S; disulfide: S-S; isocyanate: N=C=O} \\
Cyclic siloxanes & thioester; isocyanate &
\texttt{thioester: [CX3](=O)S; isocyanate: N=C=O} \\
Cyclic silazanes & carbonate; thioester; disulfide; isocyanate &
\texttt{carbonate: [CX3](=O)O[OX2]; thioester: [CX3](=O)S; disulfide: S-S; isocyanate: N=C=O} \\
NTA & thioester; isocyanate &
\texttt{thioester: [CX3](=O)S; isocyanate: N=C=O} \\
\hline
\end{tabular}
\caption{SMARTS patterns used in Rule 3 to exclude side-chain motifs that may interfere with specific ROP monomer classes.}
\label{tab:rule3}
\end{table}

\begin{table}[H]
\centering
\begin{tabular}{lll}
\hline
\textbf{Feature} & \textbf{Keep / Exclude} & \textbf{Examples} \\
\hline
ROP rings & Keep &
lactone, lactam, cyclic carbonate, cyclic ether, cyclic sulfide,\\
&& 1,2-dithiolane, cyclic phosphoester, cyclic siloxane,\\
&& cyclic silazane, NCA, NTA, OCA, epoxide,\\
&& cycloolefin, CKA \\

Only non-ROP rings & Exclude &
cycloalkanes, cycloketones, aromatic rings \\

Low-strain special cases & Exclude &
$\gamma$-thiolactone-like rings, tetrahydropyran-like rings \\
\hline
\end{tabular}
\caption{Patterns used in Rule 4 to exclude molecules lacking an ROP motif or containing low-strain ring systems that are unlikely to undergo ring-opening.}
\label{tab:rule4}
\end{table}

\end{document}